\documentclass[12pt]{article}
\pdfoutput=1

\usepackage{epsfig}
\begin{document}
\begin{center}
{\Large \bf Automatic Step Size Selection \\ 
in Random Walk Metropolis Algorithms}
\end{center}

\begin{center}
Todd L. Graves\\
Statistical Sciences\\
Los Alamos National Laboratory\\
Los Alamos, NM 87545\\
LA-UR 11-01936
\end{center}

\begin{center}
{\bf Abstract}
\end{center}

Practitioners of Markov chain Monte Carlo (MCMC) may hesitate to use
random walk Metropolis--Hastings algorithms, especially
variable-at-a-time algorithms with many parameters, because these
algorithms require users to select values of tuning parameters (step
sizes).  These algorithms perform poorly if the step sizes are set to
be too low or too high.  We show in this paper that it is not
difficult for an algorithm to tune these step sizes automatically to
obtain a desired acceptance probability, since the logit of the
acceptance probability is very nearly linear in the log of the step
size, with known slope coefficient.  These ideas work in most applications, 
including single parameter or block moves on the linear, log, or logit scales. 
We discuss the implementation of this algorithm in the software package YADAS.
\\

\noindent

\section{Introduction}
Some users of Markov chain Monte Carlo (MCMC) methods in Bayesian
statistical analyses may be reluctant to implement samplers that rely
on random walk Metropolis-Hastings moves, because these samplers
require users to choose values of tuning parameters.  The performance
of the samplers are sensitive to the choices of these parameters.  As
a result, users may elect to modify the statistical model they use to
allow them to use the Gibbs sampler with tractable conditional
distributions.  This fear of tuning parameters is unfounded.  In the
vast majority of examples, it is very easy for an algorithm to tune
these parameters automatically in an initial burn-in phase.  This is
true because it generally suffices to aim for a given acceptance rate
for proposed moves, and because the logit of the acceptance rate is
almost always linear in the step size (proposal standard deviation),
with a slope coefficient that can be taken to be known.  The burn-in phase 
can feature a designed experiment to estimate the intercept of the logistic 
regression, which can then be used to set the values of the tuning parameters 
for the remainder of the algorithm.  This algorithm is implemented in 
YADAS (Graves, 2001, 2003, 2007).

Interesting cases will involve the need to tune several step sizes
simultaneously.  There is as yet no indication that this should pose
any problems.  Critically, the approach and the software
implementation work not just for single parameter updates but also for
YADAS's block updates, for positive parameters updated on the log scale, and 
for probability parameters updated on the logit scale.  

In a block update, a Metropolis(-Hastings)
move proposes a simultaneous alteration to multiple parameters, and
such updates are very helpful in situations with highly correlated
parameters.  For example, suppose that in a hierarchical model,
several \textit{a priori} exchangeable parameters $\theta_i$ have a
common unknown mean $\mu$, and the remaining parameters are such that
the value of $\mu$ is uncertain, but the $\theta_i$ are all close to
$\mu$.  (This situation applies in one-way ANOVA with some values for
the random effect variances; see Graves (2003b), Example 4.)  Moving
$\mu$ and the $\theta_i$ independently works poorly, because none of
these parameters can move quickly due to being forced to remain close
to the others.  However, excellent mixing is possible when a naive
algorithm is augmented with a step that chooses $Z \sim N(0,1)$, and
proposes candidate values $\theta_i^C$ and $\mu^C$ according to
$\theta_i^C = \theta_i + sZ$ for all $i$ and $\mu^C = \mu + sZ$.
Choosing a relatively efficient value of $s>0$ is the topic of this
paper.  See Graves, Speckman, and Sun (2003), for example, for theory
and examples of block updates.

This algorithm is not a panacea, since many samplers have more serious
problems than the choice of tuning parameters.  However, this
algorithm can play a key role in problems which only need suitable
values of step sizes, even a large number of them.

\subsection*{Related work}
Gelman, Roberts, and Gilks (1995) work with algorithms consisting 
of a single Metropolis move (not variable-at-a-time), and obtain 
many interesting results for the $d$-dimensional spherical multivariate 
normal problem with a symmetric proposal distributions, including that 
the optimal scale is approximately $2.4d^{-1/2}$ times the scale of 
the target distribution, which implies optimal acceptance rates of 
0.44 for $d=1$ and 0.23 for $d \rightarrow \infty$.  

Roberts and Rosenthal (2001) evaluate scalings that are optimal (in
the sense of integrated autocorrelation times) asymptotically in the
number of components.  They find that an acceptance rate of 0.234 is
optimal in many random walk Metropolis situations, but their studies
are also restricted to algorithms that consist of only a single step in
each iteration so are not directly applicable here, and in any case
they conclude that acceptance rates between 0.15 and 0.5 do not cost
much efficiency.

Yeung and Wilkinson (2002) model lagged autocorrelation as, for
example, a quadratic response surface in tuning parameters, and use
stochastic search algorithms to obtain good values for the tuning
parameters.  Their focus is on comparing various MCMC algorithms 
(standard Gibbs sampling vs. block updating, for example) more than 
on demonstrating the efficiency of the tuning methodology.  
Pasarica and Gelman (2004) aim to maximize the expected squared 
distance between successive MCMC samples, since this is equivalent 
to minimizing first order autocorrelation, and use importance 
sampling estimates of this quantity for several step sizes and 
numerical optimization.  This procedure is promising and includes 
the case of simultaneously optimizing multiple tuning parameters  
Its performance is likely to suffer with increasing 
dimensionality of the tuning parameter, but it may be adaptable 
to several individual optimizations instead of a single large one, 
which would be appropriate for the variable-at-a-time case.  

Andrieu and Thoms (2008) present several important adaptive algorithms
that tune step sizes and create block updates, among other things.  We
encourage the reader to investigate their work as well.

\section{Philosophy}
Markov chain Monte Carlo (MCMC) algorithms are often too badly flawed
to be saved by judicious choice of step sizes.  This work assumes that
they are not and that the algorithm contains the right composition of
steps.  If correlations between parameters or multimodality prevent
adequate exploration of the posterior distribution regardless of the
values of tuning parameters, more drastic measures are necessary.
See, for example, Liu and Rubin
(2002), and Andrieu and Thoms (2008). 

We aim to tune step sizes to achieve a desired acceptance rate that is
the same for every update of every problem (we use $1/e$ as the target
acceptance rate, because this gives the false impression that the
target is a result of a theoretical study).  This strategy is almost
certainly suboptimal but has worked well in practice for us. Sherlock,
Fearnhead, and Roberts (2010) also tune various algorithms to attain
target acceptance rates, and their Algorithm 2 tunes step sizes of
univariate updates to attain acceptance rates between 0.4 and 0.45.

Once the goal of tuning step sizes to attain acceptance rates of $1/e$
(say) is accepted, it is surprisingly easy to achieve it.  We find
that the logit of the acceptance rate is very nearly linear in the log
of the step size.  Consider, for example, the case where the posterior
distribution $f(x) = (2\pi)^{-1/2} \exp(-x^2/2)$ is standard normal.
Given that the current state of the Markov chain is $x$, we propose a
new state $y \sim N(x, s^2)$ (i.e.\ the transition proposal density is
$T(x,y) = (2\pi s)^{-1/2} \exp\{-(y-x)^2/2s^2\}$), define $R(x,y) =
\frac{T(y,x)}{T(x,y)}\frac{f(y)}{f(x)}$, and accept the move to state
$y$ with probability $\min\{1,R(x,y)\}$.  The long--run acceptance
rate is obtained by integrating the acceptance probability with
respect to the joint distribution of $(x,y)$, namely
\begin{eqnarray*}  & & \int_{-\infty}^{\infty} \int_{-\infty}^{\infty} \min\{1, R(x,y)\} 
T(x,y)\ dy\ f(x)\ dx \\ 
& = & \int \int_{\{(x,y): R(x,y) \leq 1\}} T(y,x)\ f(y)\ dy\ dx + 
\int \int_{\{(x,y): R(x,y) > 1\}} T(x,y)\ f(x)\ dy\ dx,
\end{eqnarray*}
which, if the proposal distribution is symmetric (if we are using 
the Metropolis algorithm, with $T(x,y) = T(y,x)$), and if it is 
also continuous, reduces to
$$2\int \int_{\{(x,y): f(y) > f(x)\}} T(x,y)\ f(x)\ dy\ dx.$$
Observing that in the case where $x$ and $y|x$ are both 
normal, $f(x) < f(y)$ if and only if $|x| > |y|$, this can be 
rewritten 
$$2 \int_{\infty}^{\infty} \int_{-|x|}^{|x|} (2\pi |\Sigma_s|)^{-1} 
\exp\left\{-\frac{1}{2}(x\ y) \Sigma_s^{-1} \left(\begin{array}{c} x \\ y\end{array}\right)\right\}\ dy\ dx,$$ 
where $\Sigma_s$ is the covariance matrix of $(x,y)$ 
($\Sigma_{11} = \Sigma_{12} = \Sigma_{21} = 1$ and $\Sigma_{22} = 1 + s^2$).  
Next change variables to the independent standard normal 
$(u\ v)^T = \Sigma_s^{-1/2}(x\ y)^T$.  It can be verified directly 
that
\[\Sigma_s^{1/2} = (2+2s+s^2)^{-1}
\left( \begin{array}{cc} 
1+s & 1 \\ 
1   &  1+s+s^2 \end{array} \right), \]
so that 
$$\{(u,v): |x|>|y|\} = \{(u,v): |(1+s)u + v| > |u + (1+s+s^2)v|\}.$$
Our acceptance rate therefore reduces to 
\begin{eqnarray*} 
 & & 2 \int\int_{\{(u,v):|(1+s)u + v| > |u + (1+s+s^2)v|\}} (2\pi)^{-1}
\exp\left\{-\frac{1}{2}(u^2+v^2)\right\}\ dv\ du \\ 
 & = & 4 \int_0^\infty \int_{-u\frac{2+s}{2+s+s^2}}^{u\frac{1}{1+s}} 
(2\pi)^{-1} \exp\left\{-\frac{1}{2}(u^2+v^2)\right\}\ dv\ du \\ 
 & = & 4 \int_0^\infty \left\{\Phi\left(\frac{x}{1+s}\right)-\Phi\left(\frac{-x(2+s)}{2+s+s^2}\right)\right\} \phi(x)\ dx. 
\end{eqnarray*}
Gelman, Roberts, and Gilks (1995) report, without elaboration, 
that this acceptance rate 
``can be determined analytically'' and equals 
$\frac{2}{\pi}\arctan(\frac{2}{s})$.  One way to 
check this is to substitute $t = 2/s$, and show that the derivative 
of this function with respect to $t$ is equal to $\frac{2}{\pi}(1+t^2)^{-1}$.
After differentiating, the integration can easily be done in closed 
form, leaving some algebra to be done.  

\begin{figure}
\begin{center}
\includegraphics[scale=0.4]{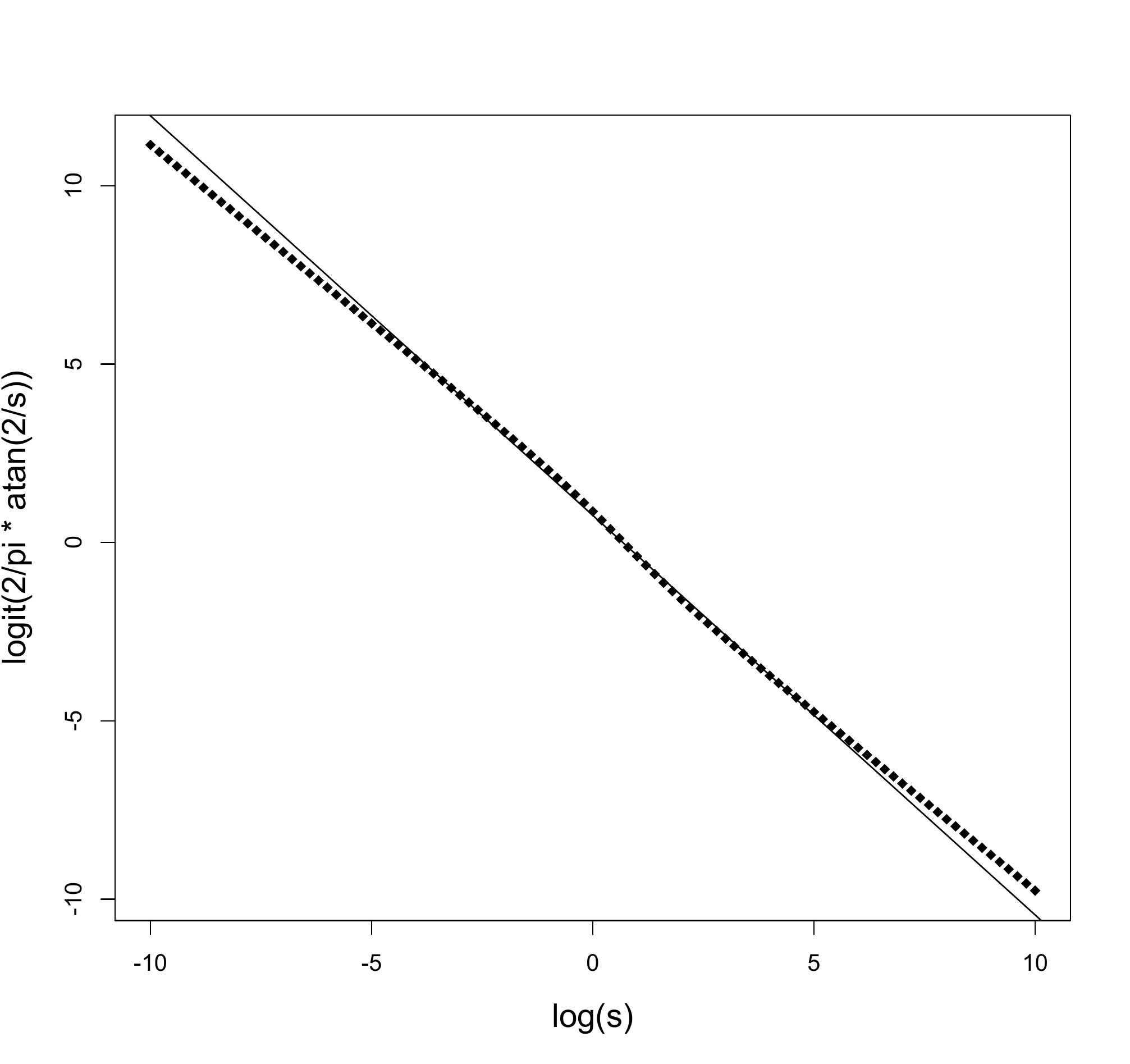}
\caption{Approximate linear relationship of the logit acceptance rate and the log of the step size.}
\label{fig:linear}
\end{center}
\end{figure}

Plotting the logit of $\frac{2}{\pi}\arctan(\frac{2}{s})$ against
$\log(s)$ demonstrates the very close approximate linearity; see
Figure~\ref{fig:linear}.  Denote this function by $p_1(s)$; the
subscript $1$ refers to the standard deviation of the posterior
distribution.  We have $$ \mbox{logit}\ p_1(s) \simeq 0.76 - 1.12
\log(s)$$ using a least squares fit of the numerically integrated
function.  Observe further that, since clearly $p_\sigma(s) =
p_1(s/\sigma)$, we have
$$ \mbox{logit}\ p_\sigma(s) \simeq 0.76 - 1.12 \log (s/\sigma) = (0.76
+ 1.12 \log \sigma) - 1.12 \log s, $$ so that the slope of the
relationship is nearly equal to the constant $-1.12$, independently of
$\sigma$.  Clearly, this result holds for other means for $x$, and one
expects the same result to hold where $x$ is a single component of a
parameter with a(n approximately) multivariate normal posterior.
Therefore, to find an appropriate step size for an approximately 
normal marginal distribution, one can consider collecting acceptance 
rate data for various step sizes and fitting a logistic regression 
with known slope.  The marginal distribution does not need to be 
very close to normal, either: for example, simulations indicate that 
the slope for an exponential posterior distribution is about $-1.08$, and so 
is the slope for a $t$ distribution with $2$ degrees of freedom.  
These slopes are close enough to $-1.12$ that it is better to use 
the fixed slope than to try to estimate a slightly different one 
using data.  

Certainly, one could use the actual arctangent relationship to try to
choose a good $s$: in the univariate example, if $p$ is the desired
acceptance rate, then we obtain $s = 2\sigma/\tan(\frac{\pi}{2}p)$,
where $\sigma$ is the posterior standard deviation, so we only need to
estimate $\sigma$.  (Examples of results include $s=2.4\sigma$ for
$p=0.44$, $s=3.1\sigma$ for $p = 1/e$, and $s=5.3\sigma$ for
$p=0.23$.)  However, in variable-at-a-time random walk Metropolis
updates, one expects that the proper interpretation of $\sigma$ is not
the posterior standard deviation but the average conditional standard
deviation, which is presumably more difficult to estimate from a
Metropolis algorithm.

In our experience, this approach works equally well 
for positive parameters updated on the log scale with 
\[T(x,y) = (2\pi s)^{-1/2} y^{-1} \exp\{-(\log y-\log x)^2/2s^2\}\]
and with parameters in $(0,1)$ updated on the logit scale with
\[T(x,y) = (2\pi s)^{-1/2} [y(1-y)]^{-1} \exp\{-(\mbox{logit } y-
\mbox{logit } x)^2/2s^2\}.\] We essentially always update standard
deviation parameters, and, respectively, probability parameters in
these ways.  Our approach also works well if the parameter $x$ is a vector
of probabilities that sum to one, and if we propose a new value by
adding a Gaussian perturbation to $\mbox{logit } x_i$ and rescaling
the other $x_j$'s so that they still sum to one (and we have a step
like this for each $i$).

\subsection{Trial stage and logistic regression}
To take advantage of this relationship, we begin our MCMC algorithm
with a trial stage: the user specifies an initial guess for each
proposal standard deviation.  The trial stage loops
through thirteen (for example) logarithmically spaced step sizes fifty 
times for each, and monitors the number of accepted moves for each 
step size.  If one then overlooks the fact that the slope in this 
regression can be taken to be known, one can then fit a logistic regression: 
$$ \mbox{logit}(\mbox{acceptance probability with step size }s) = 
	a + b \log(s)$$
using Newton-Raphson, and 
use the estimated parameters to try to hit a target acceptance rate 
$\hat s = \hat b^{-1} (\mbox{logit}(p_{target}) - \hat a)$.  
The Newton-Raphson algorithm works as follows.  
Let $\hat a^{(k)}$ and $\hat b^{(k)}$ be the estimates of $a$ and $b$ after 
$k$ iterations.  Let $n_i$ be the number of trials for the 
$i$th step size, $x_i$ be the number of accepted moves, and 
$s_i$ be the $i$th step size.  
Let $\hat p_i^{(k)} = \mbox{logit}^{-1} (\hat a^{(k)} + \hat b^{(k)}\log s_i)$
be the estimated acceptance probabilities for step size $i$ after 
$k$ iterations.  
Further let $A^{(k)} = \sum_i n_i \hat p_i^{(k)} (1- \hat p_i^{(k)})$,
$B^{(k)} = \sum_i n_i (\log s_i) \hat p_i^{(k)} (1- \hat p_i^{(k)})$, and
$C^{(k)} = \sum_i n_i (\log s_i)^2 \hat p_i^{(k)} (1- \hat p_i^{(k)})$. 
The Newton--Raphson equations are 
{ \small
\begin{eqnarray*}
\hat a^{(k+1)} & = & a^{(k)} + \{A^{(k)} \sum_i (x_i - n_i \hat p_i^{(k)}) - 
 B^{(k)} \sum_i (\log s_i)(x_i - n_i \hat p_i^{(k)})\}/(A^{(k)}C^{(k)} - B^{(k)2}) \\
\hat b^{(k+1)} & = & b^{(k)} - \{B^{(k)} \sum_i (x_i - n_i \hat p_i^{(k)}) + 
 C^{(k)} \sum_i (\log s_i)(x_i - n_i \hat p_i^{(k)})\}/(A^{(k)}C^{(k)} - B^{(k)2}).\\
\end{eqnarray*}
}
In practice, this algorithm converges quickly (in less than twenty 
iterations) with starting values $a^{(0)} + b^{(0)} = 0$, for 
essentially all the data we have tried.  

This algorithm has been implemented in YADAS in case we find an
application where the logistic slope differs substantially from -1.12.
Most often, though, one should fix the slope and estimate the
intercept alone.  Using the notation above, the algorithm then updates
the intercept $a$ as follows:
$$ a^{(k+1)} = a^{(k)} + \frac{\sum_i n_i (\frac{x_i}{n_i} - \hat
p_i^{(k)})} {\sum_i n_i \hat p_i^{(k)}(1-\hat p_i^{(k)})}. $$ This
algorithm is less numerically stable.  It can diverge readily, even
when the algorithm is augmented by step halving when the change to $a$
shrinks the likelihood.  For this and other reasons, we actually
perform the estimate with a ``prior'' for $a$ included for
regularization purposes.  A normal prior with mean $\mu_a = -3$ and
standard deviation $\sigma_a = 5$ implies that the optimal step size
will normally be between $10^{-5}$ and $10^3$, and modifies the
updating algorithm for $a$ to
$$ a^{(k+1)} = a^{(k)} + 
\frac{\sum_i n_i(\frac{x_i}{n_i} - \hat p_i^{(k)}) - \frac{a^{(k)} - \mu_a}
{\sigma_a^2}}
{\sum_i \hat n_i \hat p_i^{(k)}(1-\hat p_i^{(k)}) + \frac{1}{\sigma_a^2}}. $$

One may note that an experiment with thirteen different step sizes may
be very conservative compared to what is actually required.  In
principle, since the slope is known, if one knows the acceptance rate
accurately for one step size, one can use that information alone to
tune it to a different acceptance rate.  However, spending a few
hundred iterations in this manner has not seemed overly expensive to
us, since we normally run the algorithm for many more iterations after
tuning.

\subsection{Simulation experiment}
Here we discuss a simulation experiment that explores how accurate the
initial guesses for step sizes need to be, and how many
logarithmically spaced step sizes and how many attempted steps at each
step size are needed.  Our experiment assumed that the logistic
regression model was correct, and that the acceptance rate for step
size $s$ is $\mbox{logit}^{-1}\{-5.7 - 1.12\log(s)\}$; the constants
were chosen so that the ``optimal'' step size is 0.01.  The full
factorial simulation experiment tried initial guesses of $0.01 \times
2^k$ for $k = \{-7, -6, \ldots, 0, \ldots, 6,7\}$, $3,5,7,9,11,13,$
and $15$ different step sizes per run, and 10,20,30,40, and 50
attempted steps per step size.  (For example, suppose that the initial
guess was 0.16 and the number of step sizes was 9.  The step sizes
considered would be $0.01, 0.02, 0.04, 0.08, 0.16, 0.32, 0.64,
1.28,$ and $2.56$.  These step sizes yield acceptance rates of
$0.368, 0.700, 0.903, 0.974, 0.993, 0.998, 1.000, 1.000,$ and
$1.000$.)  For each combination of these variables, we constructed 100
simulated data sets using the logistic regression model, estimated $a$
with fixed $b=-1.12145$ using the Newton--Raphson algorithm,
recommended a step size based on the estimated $a$ and $b$, and
counted how many times out of 100 the recommended step size would
generate an acceptance rate between $0.25$ and $0.45$.  (For the
example with initial guess 0.16 and nine step sizes, the number of
``successes'' with 10, 20, 30, 40, and 50 attempts per step size were
73, 88, 84, 92, and 91 respectively.)  The simulation results are a 
little erratic to report: the discreteness of the possibilities means 
that by chance a good step size can be chosen although the data are 
inadequate.  For example, if the initial guess is $0.01 \times 2^7$, 
we only attempt three step sizes ten times each, and we get zero 
acceptances at each step size (which happens about $92\%$ of the time), 
the algorithm will choose a step size of 0.011, for an acceptance 
rate of $34\%$.  The designs adjacent to this case all have essentially 
zero probability of getting an acceptable step size.  Less severe 
nonmonotonicities exist as well.  

We assume that the initial guess at the step size is off by a given 
power of two, and report the total trial sample size that yields a 
success rate (acceptance rate between 25 and $45\%$) of at least $95\%$.  
If one's initial guess is exactly right, a total sample size of 120-180 
is adequate, and most efficient is 40 trials at each of three step sizes.  
Not surprisingly, this minimal number of step sizes ceases to be 
efficient quickly when one's initial guess is off.  When the initial 
guess is too large by a factor of two, roughly 200 total trials are 
needed, and there are several equally efficient ways of getting there: 
20 each at 9 or 11 levels, 30 each at 7 levels, or 40 at 5 levels.  
When the initial guess is too high by a factor of 4, 8, 16, or 32, 
20 trials at each level are appropriate, and the number of levels 
should be 11, 11, 13, and 15 respectively: the number of levels needs 
to be large enough so that at least two step sizes smaller than the 
optimal are attempted.  Overestimating the step size by a factor of 
64 can be overcome with 30 trials at each of 15 levels, and an 
overestimate by a factor of 128 should be avoided.  

The situation is not symmetric when one underestimates the optimal 
step size: in fact, it is better to underestimate it a bit, which 
is unsurprising since lower step sizes imply acceptance rates of 
closer to $0.5$, making each trial more informative.  Forty trials 
at each of three levels continues to work well if the initial guess 
is too low by a factor of two or four.  Underestimates of factors 
of 8, 16, and 32 require total sample sizes of about 180, 220, and 
280 respectively, and any choice 20, 30, or 40 per step size is 
about equally effective.  

Of course, one may desire a lower or higher than $95\%$ probability 
of getting an acceptance rate between $25\%$ and $45\%$.  YADAS uses thirteen levels and sample sizes of 50 at each level, and this leads to zero failures in practice even when there are hundreds of parameters. YADAS also sends 
the chosen step sizes to output files.

\section{Implementation}
In this section we discuss the YADAS classes that can be used to 
tune step sizes.  For an introduction to YADAS's software design, 
please see Graves (2001, 2003, 2007).  

In YADAS, the definition of an algorithm is a collection of objects
implementing the \texttt{MCMCUpdate} interface.  YADAS loops through this
collection of updates, calls the \texttt{update()} method of each, and each
of these attempts to change the values of one or more unknown
parameters.  One complete cycle through the collection of updates is
one iteration in the MCMC algorithm, and the current values of the
parameters are then sent to output files.  The simplest example of a
object implementing the \texttt{MCMCUpdate} interface is a parameter.  A
parameter's \texttt{update()} method loops through the components in the
parameter, attempting a Gaussian random walk Metropolis move to each.
Another type of update is the \texttt{MultipleParameterUpdate}, which has
the capability of proposing a Metropolis or Metropolis--Hastings move
that affects multiple parameters simultaneously: the proposed move is
defined using a \texttt{Perturber}.

\subsection{UpdateTuner}
To introduce step size tuning to YADAS, we built a class called 
\texttt{UpdateTuner} that itself implements \texttt{MCMCUpdate} and can 
be inserted into the collection of updates in a YADAS algorithm.  
The \texttt{UpdateTuner} supervises another update, altering its step 
sizes and monitoring its acceptance rates during a trial phase, 
analyzing the results of the trial experiment, and then selecting 
step sizes for the final phase of the MCMC algorithm.  To define 
an \texttt{UpdateTuner}, one specifies:
\begin{itemize}
\item An object implementing the \texttt{TunableMCMCUpdate} interface, 
which will be described below: this is the update step whose step 
size we are trying to tune.  A \texttt{TunableMCMCParameter} is an example 
of such an object;
\item an array of initial guesses for step sizes, one for each step in 
the update;
\item an integer defining the number of trial step sizes to use in the 
experiment;
\item an integer defining the number of attempts per trial step size; and
\item a target acceptance rate (we typically use $1/e$).  
\end{itemize}
The last five of these arguments can be omitted; they have default 
values of: the step sizes assigned to the update, 13, 50, 1, and $1/e$.  

\subsection{TunableMCMCUpdate}
The \texttt{TunableMCMCUpdate} interface extends the \texttt{MCMCUpdate} interface 
in the following way.  It introduces several new methods:
\begin{itemize}
\item \texttt{getStepSizes()} returns a vector of doubles, the 
values of the step sizes currently used by the update.  This is 
actually not used by the \texttt{UpdateTuner} class so may disappear.  
\item \texttt{setStepSize} has two signatures; one changes a single step size, 
the other the entire vector.  
\item \texttt{acceptances()} returns a vector of numbers of acceptances, one 
for each step size.  
\item \texttt{tuneoutput()} writes the ultimately selected step sizes to a file 
with extension \texttt{.tun}.  
\end{itemize}
If a user wants to write a new type of update and wants it to be tunable, 
the user must ensure that the new update includes appropriate definitions 
for all these methods.  Important update classes that are tunable are 
\texttt{MCMCParameter} and, unsurprisingly, \texttt{TunableMultipleParameterUpdate}.  

\subsubsection{Additions to MCMCParameter}
We included all the tuning code in the \texttt{MCMCParameter} class itself 
rather our first intention, which was to have a \texttt{TunableMCMCParameter} 
subclass.  The \texttt{TunableMCMCUpdate} methods do reasonable things.  
An \texttt{MCMCParameter} includes a vector of unknown scalar parameters, 
each of which has a step size, and these are the step sizes that are 
accessed.  

Defining parameters in a YADAS application with tuning is identical to 
applications without, the only exception being that the step sizes 
included in the definition of a parameter are initial guesses only.  

\subsubsection{TunableMultipleParameterUpdate}
In YADAS, one may add steps to an MCMC algorithm that (attempt to) improve 
mixing by moving multiple parameters together.  This functionality 
is centered in the \texttt{MultipleParameterUpdate} class, and especially the 
\texttt{Perturber} interface.  A \texttt{MultipleParameterUpdate} consists only of 
an array of parameters and a \texttt{Perturber}.  The \texttt{Perturber} includes 
all the specialization, such as the method \texttt{perturb()},
which produces proposed new values of the parameters, given their old 
values, and also calculates the ratio of proposal probabilities that 
appears in the acceptance rate for Metropolis--Hastings moves.  
The \texttt{perturb()} method frequently depends on one or more tunable step 
sizes.  A canonical example is the \texttt{NewAddCommonPerturber}.  It is quite 
common that a posterior distribution is approximately a function of 
differences of some parameters, so that the posterior is relatively 
insensitive to transformations that add a common constant to all those 
parameters.  This is what \texttt{NewAddCommonPerturber} does: it samples a 
random Gaussian $Z$ with some standard deviation $s$, and proposes a 
Metropolis move in which several parameters are incremented by $Z$.  
More generally, the parameters can be divided into groups, each of 
which gets its own random $Z_j$ and each of which has its own 
standard deviation $s_j$.  In this case we want to tune the $s_j$'s.  

All the \texttt{Perturber}s included in the YADAS package are tunable.  
We did not include the tuning capability in the \texttt{Perturber} interface itself
(rather it is in the \texttt{TunablePerturber} interface) because we didn't 
want to make it more difficult than necessary for users to write new 
\texttt{Perturber}s, but we will try to ensure that the ones we write are 
as usable as possible, and that includes making them tunable.  

YADAS also includes a class \texttt{ReversibleJumpUpdate} which is not yet 
tunable; further study is required before it is clear that tuning 
acceptance rates is appropriate in reversible jump problems.  

\section{Examples}
Finally, in this section, we present some examples of the tuning 
process in practice.  First, we work with a normal example with
unknown mean and variance and tune both step sizes.  Second, 
we work with a one-way ANOVA that has been parameterized poorly with 
resultant poor mixing, fix the mixing with a \texttt{MultipleParameterUpdate}, 
and tune its step size along with the others in the problem.  
The source code and data for these examples are available on the 
YADAS website \texttt{yadas.lanl.gov}; follow the ``Download'' link and 
then download the zipped directory of examples.  

\subsection{Normal example}
In this problem, Example 10 on the YADAS web site, we have data $y_i
\sim N(\mu, \sigma^2)$ for $i= 1,\ldots,N$, with priors $\mu \sim
N(a_\mu, b_\mu^2)$ and $\sigma \sim \Gamma(a_\sigma, b_\sigma)$
(according to our parameterization, $\sigma$ has prior mean $a_\sigma
b_\sigma$).  Each iteration in the MCMC algorithm has two steps: a
Metropolis move in which we propose a Gaussian random walk move to
$\mu$ (i.e.\ $\mu^\prime = \mu + s_1 Z_1$, where $Z_1 \sim N(0,1)$,
and a Metropolis--Hastings move in which we propose a lognormal
adjustment to $\sigma$ (i.e.\ $\sigma^\prime = \sigma \exp(s_2 Z_2)$, 
where $Z_2 \sim N(0,1)$.  $s_1$ and $s_2$ must be tuned.  

The key piece of code in the tuning application is 
\begin{verbatim}
MCMCUpdate[] updatearray = new MCMCUpdate[] 
{ new UpdateTuner(mu, d0.r("mumss"), 13, 50, 1, Math.exp(-1)), 
  new UpdateTuner(sigma, d0.r("sigmamss"), 13, 50, 1, Math.exp(-1)) };
\end{verbatim}
in which we define the algorithm to consist of two update steps as 
described above, and whose step sizes will be tuned.  Beginning with the 
first \texttt{UpdateTuner}, \texttt{mu} is the definition of the update step (here, 
a Gaussian random walk update to $\mu$).  The expression 
\texttt{d0.r(``mumss'')} defines a vector of initial guesses for step sizes 
(only one here, and this expression gets them from an input file).  
The 13 refers to the number of different step sizes to experiment 
with, the experiment will attempt 50 moves for each step size, the 1 
means only one cycle of experimentation, and the last argument 
implies that the step size will be tuned for an acceptance rate of 
$1/e$.  The fact that $\sigma$ will be updated using Gaussian random 
walk moves on the $\log$ scale has been determined elsewhere 
(\texttt{sigma} was defined to be a \texttt{MultiplicativeMCMCParameter} instead 
of just a \texttt{MCMCParameter}).

\subsection{Badly parameterized one-way ANOVA example}
\begin{figure}
\begin{center}
\includegraphics[scale=0.5]{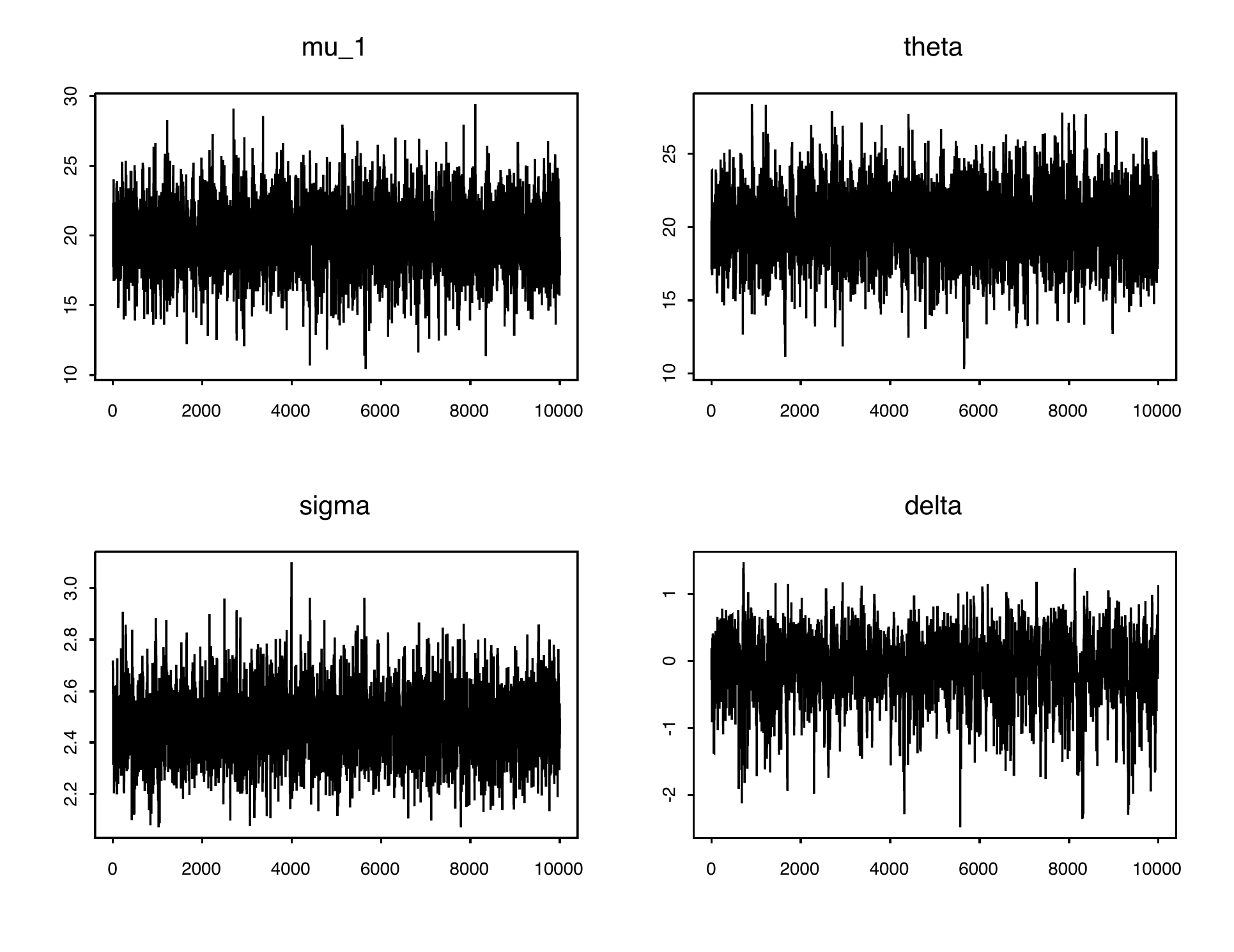}
\caption{Trace plots for $\mu_1, \theta, \log \sigma,$ and $\log \delta$ in 
the one--way ANOVA example.}
\label{fig:anova}
\end{center}
\end{figure}

This example, Example 11 on the YADAS website, shows that the tuning
procedure works even in cases where multiple parameters are updated at
once.  This is a one-way analysis of variance example that is commonly
used to illustrate mixing difficulties in MCMC algorithms that can be
solved with reparameterization.  Data $y_{ij}$ are normal with means
$\mu_i$ and common standard deviation $\sigma$ for $i = 1,2,\ldots,I$
and $j = 1,2,\ldots,n_i$.  Here we have taken the $\mu_i$ to have a
$N(\theta, \delta^2)$ prior, $\theta$ has a flat hyperprior, and
$\sigma$ and $\delta$ have Gamma priors.  This parameterization works
fine except when $\delta$ is too small compared to $\sigma$ (the
sample sizes also drive what is meant by ``too small'').  In this
case, the $\mu_i$ and $\theta$ have high posterior correlation so that
it works poorly to update them individually.  Many solutions exist
including reparameterization or block Gibbs updates, but here we use a
\texttt{MultipleParameterUpdate} that augments the standard variable-at-a-time 
Gaussian random walk Metropolis with an additional step that proposes 
adding a common random Gaussian perturbation to all the $\mu_i$ and $\theta$.  
The posterior distribution is relatively invariant to moves like this, 
at least when $\delta$ is small, as it is in the supplied input files.  
The code to define the update algorithm is as follows:
{\small
\begin{verbatim}
MCMCUpdate[] updatearray = new MCMCUpdate[] { 
    new UpdateTuner (mu), new UpdateTuner (theta), 
    new UpdateTuner (sigma), new UpdateTuner (delta), 
    new UpdateTuner ( new TunableMultipleParameterUpdate 
        ( new MCMCParameter[] {mu, theta},
          new NewAddCommonPerturber ( new int[][] { d2.i(0), d0.i(0) }, 
                                      d0.r("mtmss") ), direc + "mtu") ) };
\end{verbatim}
}
Here we have used the default values for the experimental design 
descriptors (13 different trial step sizes for 50 attempts each, and so on).  
This example successfully tunes a total of four scalar parameters that are 
updated on the linear scale, two updated on the log scale, and one 
multiple parameter update.  Trace plots of the MCMC iterations are 
shown in Figure~\ref{fig:anova}.  

\section{Conclusions}
There are many other effective ways of tuning step sizes in MCMC
algorithms; we do not claim that our method is substantially superior
to the alternatives, but it is simple, intuitive, and versatile. We
readily admit that many problems (for example, those with posterior
distributions with curved contours or multiple separated modes)
generally feature MCMC difficulties that cannot be adequately solved
by tuning step sizes alone.  However, our method works well for even a
large number of tuning parameters and in a variety of posterior
distributions.

\begin{center}
{\bf \Large References}
\end{center}

\begin{description}
\item[] Andrieu, C.\ and Thoms, J.\ (2008).  A tutorial on adaptive MCMC. 
{\em Statistics and Computing} {\bf 18}: 343-373.  

\item[] Gelman, A.B., Carlin, J.S., Stern, H.S., and Rubin, D.B.\
(1995). {\em Bayesian Data Analysis}, Chapman and Hall/CRC, Boca Raton.

\item[] Gelman, A., Roberts, G.O., and Gilks, W.R.\ (1995).  
Efficient Metropolis jumping rules.  In 
{\em Bayesian Statistics 5} (J.M.\ Bernardo, J.\ Berger, A.P.\ Dawid, 
and A.F.M.\ Smith, eds.)  Oxford: Oxford University Press.  

\item[] Graves, T.L. (2001) ``YADAS: An Object-Oriented Framework for
Data Analysis Using Markov Chain Monte Carlo,'' Los Alamos National
Laboratory Technical Report LA-UR-01-4804.

\item[] Graves, T.L. (2003). ``An Introduction to YADAS,''
\texttt{yadas.lanl.gov}.

\item[] Graves, T.L. (2007). Design Ideas for Markov Chain Monte Carlo
  Software.{\em Journal of Computational and Graphical Statistics}
  16:24-43.

\item[] Graves, T.L., Speckman, P.L., and Sun, D.\ (2004). 
``Characterizing and eradicating autocorrelation in MCMC algorithms 
for linear models,'' Los Alamos National Laboratory Technical Report 
LA-UR-04-0486.  

\item[] Pasarica, C.\ and Gelman, A.\ (2010).
Adaptively scaling the Metropolis algorithm using expected squared 
jumped distance.   {\em Statistica Sinica} {\bf 20}, 343-364.

\item[] Roberts, G.O.\ and Rosenthal, J.S.\ (2001). 
Optimal scaling for various Metropolis--Hastings algorithms, 
{\em Statistical Science} {\bf 16}, 351--367.  

\item[] Sherlock, C., Fearnhead, P., and Roberts, G.O. (2010). 
The random walk Metropolis: linking theory and practice through a case study.
{\em Statistical Science} {\bf 25}: 172-190.  

\item[] Yeung, S.K.H.\ and Wilkinson, D.J.\ (2002). 
Adaptive Metropolis-Hastings samplers for the Bayesian analysis 
of large linear Gaussian systems. 
{\em Computing Science and Statistics} {\bf 33}: 128-138. 

\end{description}

\end{document}